\documentclass[twocolumn,showpacs,amsfonts,aps,prc,nofootinbib,floatfix]{revtex4}

\usepackage{bm}
\usepackage{graphicx}
\usepackage{amsmath}

\def\la{\langle}
\def\ra{\rangle}

\def\EP{\mathrm{EP}}

\def\pT{p_\mathrm{T}}

\begin{document}


\title{Fluctuating flow angles and anisotropic flow measurements}

\author{Ulrich Heinz}
\author{Zhi Qiu}
\author{Chun Shen}
\affiliation{Department of Physics, The Ohio State University,
  Columbus, OH 43210-1117, USA}

\begin{abstract}
Event-by-event fluctuations in the initial density distributions of the fireballs created in relativistic heavy-ion collisions lead to event-by-event fluctuations of the final anisotropic flow angles, and density inhomogeneities in the initial state cause these flow angles to vary with the transverse momentum of the emitted particles. It is shown that these effects lead to characteristically different transverse momentum dependencies for anisotropic flow coefficients extracted from different experimental methods. These differences can be used to experimentally constrain flow angle fluctuations in the final state of heavy-ion collisions which, in turn, are sensitive to the initial state density fluctuations and the shear viscosity of the expanding fireball medium.
\end{abstract}


\date{\today}

\maketitle

\section{Introduction}
\label{sec:1}

Due to quantum fluctuations of the positions of the nucleons inside the colliding nuclei, and of the positions of the colored quark and gluon constituents inside each nucleon, the density of the fireball matter created in collisions between ultra-relativistic heavy ions is highly inhomogeneous in the plane transverse to the beam direction and fluctuates from event to event, even for collisions with identical impact parameters. Experiments at the Relativistic Heavy Ion Collider (RHIC) \cite{Arsene:2004fa} and their theoretical interpretation have established that this matter quickly thermalizes into a quark-gluon plasma (QGP) \cite{Heinz:2001xi}. The initial density inhomogeneities lead to highly anisotropic pressure gradients, causing an anisotropic collective expansion of the fireball whose harmonic flow coefficients $v_n$ and associated flow angles $\Psi_n$ (both defined below) fluctuate from collision to collision \cite{Alver:2008zza}.

While $v_n$ fluctuations and the effect of their variance on different methods for measuring $v_n$ have been studied extensively over the last few years (for recent reviews and references to the original literature see \cite{Voloshin:2008dg,Heinz:2013th}), flow angle fluctuations and correlations have only recently found attention \cite{Qiu:2011iv,Teaney:2010vd,Gardim:2011xv,Teaney:2012ke,Jia:2012ma,Jia:2012sa,Qiu:2012uy,Ollitrault:2012cm,Gardim:2012im}. Gardim {\it et al.} \cite{Gardim:2012im} pointed out that, since the fluctuating flow angles $\Psi_n$ depend on transverse momentum $p_T$ and rapidity $y$, the usually assumed (and experimentally observed \cite{Alver:2010rt,Aamodt:2011by,Chatrchyan:2012wg,ATLAS:2012at}) factorization of the azimuthal oscillation amplitudes of the two-particle angular correlations into a product of single-particle flow coefficients is slightly broken even if these correlations are entirely due to collective flow. We show here that the $p_T$-dependence and fluctuating nature of the flow angles $\Psi_n$ also affects the $p_T$-dependence of the experimentally measured differential flow coefficients $v_n(p_T)$, and that it does so in different ways for different experimental methods of determining $v_n(p_T)$. 

\begin{figure*}[ht]
  \begin{center}
    \includegraphics[width=0.335\linewidth]{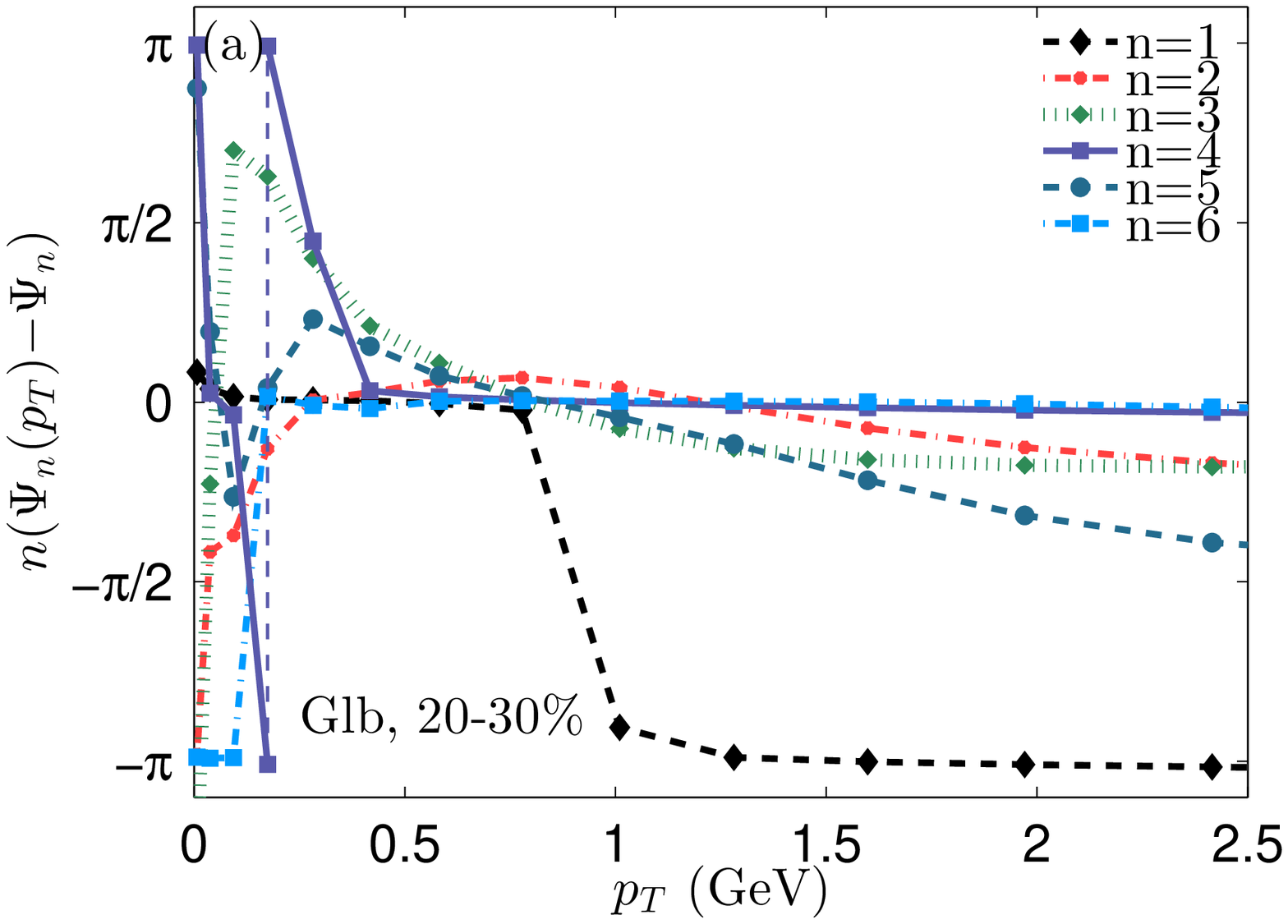}
    \includegraphics[width=0.32\linewidth]{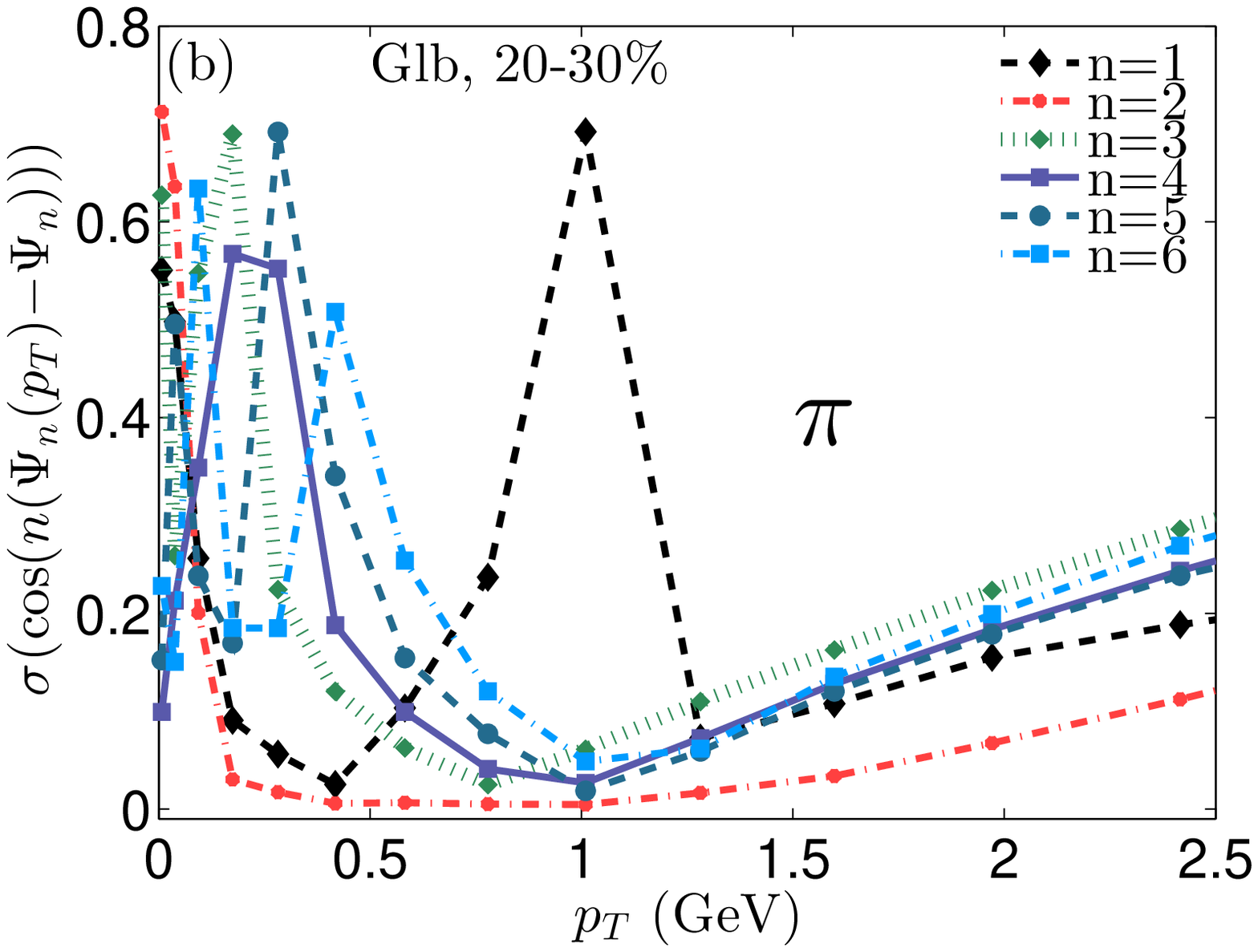}
    \includegraphics[width=0.32\linewidth]{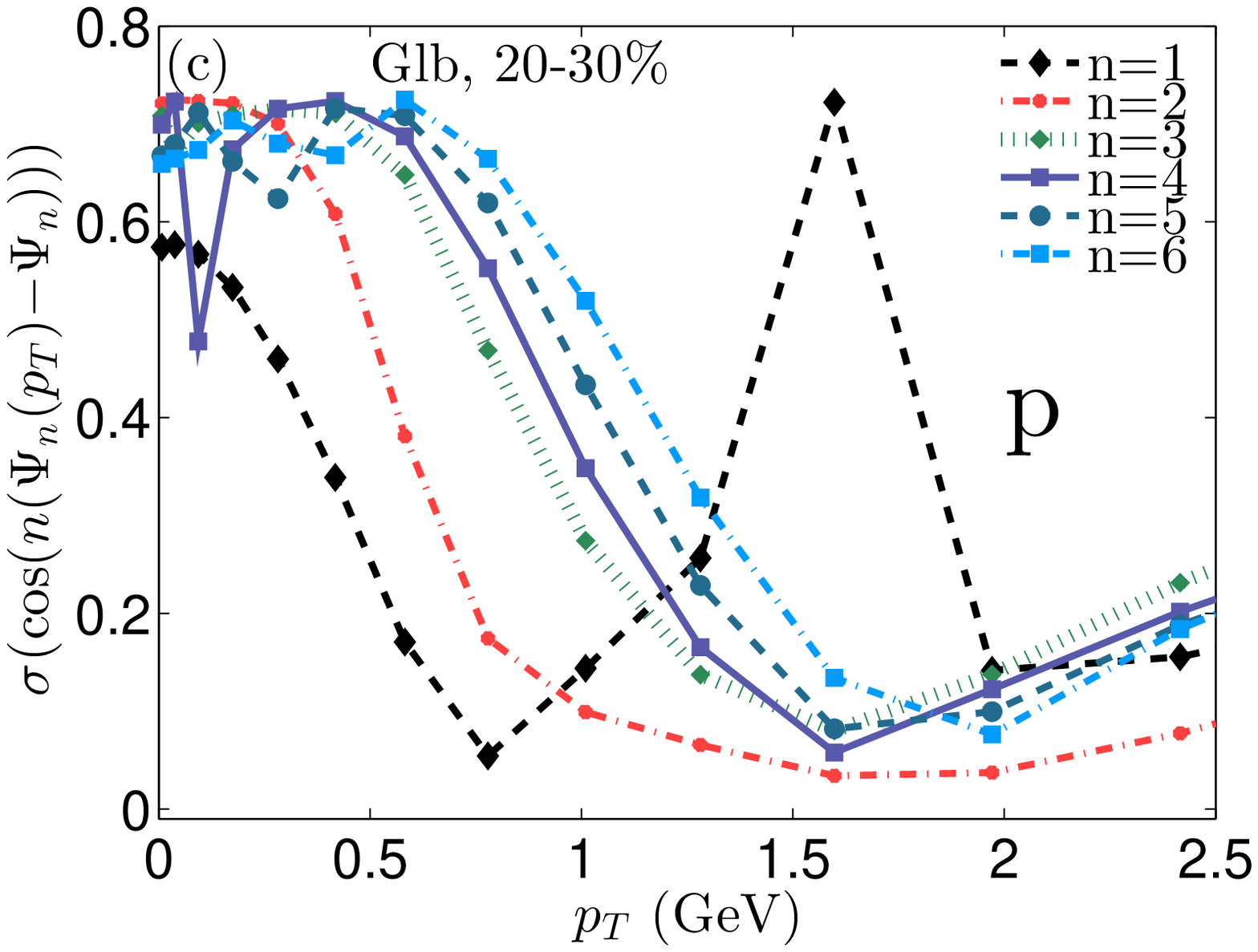}
   \end{center}
  \caption{(Color online) (a) $\pT$ dependence of the flow angles $\Psi_n(\pT)$ relative to 
    their average value $\Psi_n$ ($n{\,=\,}1,\dots,6$), for pions from a single but
    typical Pb+Pb collision event at LHC energies in the $20{-}30\%$ centrality class, computed 
    with the viscous hydrodynamic code {\tt VISH2{+}1} using an initial density profile from the 
    Monte-Carlo-Glauber model and $\eta/s{\,=\,}0.08$. Note the flip of the dipolar ($n{\,=\,}1$)  
    flow angle from 0 to $-\pi$ (happening around $\pT{\,\sim\,}0.9$\,GeV in this particular 
    event); this reflects the constraint from total transverse momentum conservation 
    \cite{Retinskaya:2012ky}. (b,c) The variance $\sigma$ of the cosine of the flow angle 
    fluctuations (which is free of ambiguities of the angles by multiples of $2\pi/n$) as a function
    of $\pT$, for pions (b) and protons (c) from 11,000 hydrodynamic events in the same 
    centrality class from which the event in (a) was taken. The variance is minimal around 
    $\pT{\,=\,}1$\,GeV for pions and around $\pT{\,=\,}1.5{-}2$\,GeV for protons (except for the 
    dipolar flow angle which flips by $\pi$ exactly in that momentum region, see panel (a)). It is
    smallest for the elliptic flow angle fluctuation $\Psi_2(\pT){-}\Psi_2$, and becomes large for 
    all flow harmonics at low $\pT$.  
    }
  \label{F1}
\end{figure*}

For each collision event the momentum distribution of finally emitted particles can be characterized by a set of harmonic flow coefficients $v_n$ and flow angles $\Psi_n$ through the complex quantities
\begin{eqnarray} 
  \label{eq1}
  V_n &=& v_n e^{in\Psi_n} 
  := \frac{\int \pT d\pT d\phi\, e^{i n \phi} \,\frac{dN}{dy \pT d\pT d\phi}}
                  {\int \pT d\pT d\phi\, \frac{dN}{dy \pT d\pT d\phi}}
  \nonumber\\
  &\equiv&\{e^{in\phi}\},
  \\
\label{eq2}
   V_n(\pT) &=& v_n(\pT) e^{in\Psi_n(\pT)} 
   := \frac{\int d\phi \, e^{i n \phi}\, \frac{dN}{dy \pT d\pT d\phi}} 
                   {\int d\phi \, \frac{dN}{dy \pT d\pT d\phi}}
   \nonumber\\
   &\equiv&\{e^{in\phi}\}_{\pT}.
\end{eqnarray}
Here $\phi$ is the azimuthal angle around the beam direction of the particle's transverse momentum $\bm{p}_\mathrm{T}$, and the curly brackets denote the average over particles from a single collision.\footnote{The average can include all charged particles or only 
    particles of a specific identified species; we will not clutter our notation to account 
    for these different possibilities.}
Eq.\,(\ref{eq1}) defines the flow coefficients and associated flow angles for the entire event, whereas Eq.\,(\ref{eq2}) is the analogous definition for the subset of particles in the event with a given magnitude of the transverse momentum $\pT$.  We suppress the dependence of both types of flow coefficients on the rapidity $y$. $v_n$ are known as the ``integrated'' anisotropic flows, $v_n(\pT)$ are called ``differential'' flows. By definition, both $v_n$ and $v_n(\pT)$ are positive definite. Hydrodynamic simulations show that in general the flow angles $\Psi_n$ depend on $\pT$, and that, as a function of $\pT$, $\Psi_n(\pT)$ wanders around the ``average angle'' $\Psi_n$ that characterizes the integrated flow $v_n$ of the entire event (see Fig.\,\ref{F1} below and also Fig.\,2 in Ref.~\cite{Ollitrault:2012cm}). Some theoretical and experimental definitions of $v_2$ have yielded values that turn negative over certain $\pT$ ranges; we will see that this is due to defining the flows of each event relative to a fixed azimuthal angle (for example, relative to the direction of the impact parameter of the collision in theoretical calculations, or relative to the ``integrated'' elliptic flow angle $\Psi_2$ in experiment), and that the same thing can happen for higher order harmonic flow coefficients when defining them relative to a fixed (i.e. $\pT$-independent) flow angle $\Psi_n$. The subject of this paper is to elucidate the origins of such differences between different anisotropic flow measures and, in particular, the manifestation of event-by-event fluctuations of the $\pT$-dependent flow coefficients $v_n(\pT)$ and flow angles $\Psi_n(\pT)$ in different experimental flow measures. 

\section{Differential flows from the event-plane method and from two-particle correlations}
\label{sec2}

The key experimental difficulty is that, due to the finite number of particles emitted in a each collision, the left hand sides of Eqs.~(\ref{eq1},\ref{eq2}) cannot be determined accurately for a single event. The $V_n$ are characterized by probability distributions that depend on the studied class of events (system size, collision energy and centrality) from which each collision takes a sample. Experimental flow measurements rely on a number of different methods that amount to taking different moments of that probability distribution, by averaging over large numbers of events. Understanding the nature of these moments and reconstructing them from theoretical event-by-event dynamical simulations are essential steps in a meaningful comparison between theory and experiment.

Our main interest lies in the event-by-event fluctuations in the initial state of the collision fireball. These are primarily caused by the finite number of nucleons (or effective collision centers) in the colliding nuclei and unrelated to detector capabilities. In addition, there are fluctuations related to the finite number of particles produced (or detected) in the event which depend on collision energy and (in part) on detector capabilities. They reflect the fact that in practice the final state of the fireball evolution, which in principle (with the appropriate dynamical evolution model) can be {\em predicted} from the initial state with perfect precision, cannot be {\em measured} with perfect precision, due to finite sampling statistics. In this paper we are not interested in the fluctuations arising from finite sampling statistics. The consequences of finite number statistical effects in a single event on the new observables proposed in this paper deserve a careful investigation that we leave for the future. Here we focus on the hydrodynamical consequences of unavoidable event-by-event fluctuations in the initial state over which we have no control since they are rooted in the internal structure of the colliding nuclei, and with which we therefore have to live in any case even after we correct the measurements for finite final state multiplicity effects.   

The most extensively used experimental methods for measuring anisotropic flows are the event-plane and two-particle correlation methods \cite{Voloshin:2008dg}. We begin with a discussion of the latter. Two-particle azimuthal correlations receive contributions from the anisotropic collective flow as well as from non-flow correlations; the latter can be minimized by appropriate experimental cuts and corrected for \cite{Voloshin:2008dg,Ollitrault:2009ie}. Again, we are not interested in non-flow correlations and will here simply ignore their existence, assuming that they have been corrected for in the experimental analysis.

Two-particle correlation measures of anisotropic flow are based on correlators of the type
\begin{equation}
\label{eq2a}
  \langle\{e^{in(\phi_1{-}\phi_2)}\}\rangle 
\end{equation}
where $\phi_1$ and $\phi_2$ are the azimuthal angles around the beam direction of two particles with transverse momenta $\bm{p}_\mathrm{T1}$ and $\bm{p}_\mathrm{T2}$, and $\langle\dots\rangle$ denotes the average over $N_\mathrm{ev}{\,\gg\,}1$ events from a set of given characteristics (e.g. of collisions in a certain centrality bin),
\begin{equation}
\label{eq4}
  \langle O \rangle = \langle\{{\cal O}\}\rangle := 
  \frac{1}{N_\mathrm{ev}}\sum_{i=1}^{N_\mathrm{ev}} \{{\cal O}\}_i,
\end{equation}      
whereas $\{\dots\}_i$ is the average of the observable ${\cal O}$ over all (or a specified subset of all) particle pairs in the event $i$:
\begin{equation}
\label{eq6}
 \{e^{in(\phi_1{-}\phi_2)}\}_i=\frac{1}{N^{(i)}_\mathrm{pairs}}\sum_{\mathrm{pairs}\in i}
   e^{in(\phi_1{-}\phi_2)}\ .
\end{equation} 
Different chosen subsets for the event-wise average $\{\dots\}_i$ define different correlation measures for the anisotropic flow coefficients as we will explain below. We note that throughout this paper we will always correlate pairs of particles of the same kind (e.g. protons with protons or charged hadrons with charged hadrons, but not protons with charged hadrons), unless specifically stated otherwise. We will also assume that they have the same rapidity $y$; generalization to particles with different rapidities is straightforward, following the procedure discussed below when we go from particles with the same to particles with different $\pT$.

The magnitudes $v_n(\pT)$ of the anisotropic flow coefficients defined in Eq.~(\ref{eq2}) fluctuate from event to event according to some probability distribution $P(v_n(\pT))$. Let us denote the rms mean of this distribution by $v_n[2](\pT){\,:=\,}\sqrt{\langle v_n^2(\pT)\rangle}$, and similarly the rms mean for the integrated flow $v_n$ by $v_n[2]{\,:=\,}\sqrt{\langle v_n^2\rangle}$. These rms means can be obtained from two-particle correlators of the type (\ref{eq2a}) as follows:
\begin{eqnarray}
\label{eq5}
  &&\!\!\!\!\!\!\!\!
  v_n^2[2](\pT) = \langle\{e^{in(\phi_1{-}\phi_2)}\}_{\pT}\rangle 
                        = \langle\{e^{in\phi_1}\}_{\pT} \{e^{-in\phi_2}\}_{\pT}\rangle ,
  \nonumber\\
  &&\!\!\!\!\!\!\!\!
  v_n^2[2] = \langle\{e^{in(\phi_1{-}\phi_2)}\}\rangle 
                = \langle\{e^{in\phi_1}\} \{e^{-in\phi_2}\}\rangle.
\end{eqnarray}
Note that for the differential flow in the first line of equation (\ref{eq5}), both particles are taken from the same $\pT$ bin, and that the event-wise pair averages $\{e^{in(\phi_1{-}\phi_2)}\}$ factorize in each event due to our assumptions (absence of non-flow two-particle correlations, independent hydrodynamic emission of particles 1 and 2). Due to $1\leftrightarrow2$ symmetry under particle exchange, the exponential can be replaced by the cosine, and we get
\begin{eqnarray}
\label{eq7}
  &&\!\!\!\!\!\!
  v_n^2[2](\pT) =
  \nonumber\\
  &&\!\!\!\!\!\!
  \left\langle\frac
  {\int d\Delta\phi \,  \cos(n\Delta\phi)\left. \frac{dN_\mathrm{pairs}}
                                  {dy_1 dy_2 p_\mathrm{T1} dp_\mathrm{T1} p_\mathrm{T2} 
                                  dp_\mathrm{T2} d\Delta\phi}\right|_{p_\mathrm{T1}{=}p_\mathrm{T2}} 
          } 
  {\int d\Delta\phi \left.\frac{dN_\mathrm{pairs}}
                                  {dy_1 dy_2 p_\mathrm{T1} dp_\mathrm{T1} p_\mathrm{T2}
                                  dp_\mathrm{T2} d\Delta\phi}\right|_{p_\mathrm{T1}{=}p_\mathrm{T2}}
          }\right\rangle,
  \nonumber\\
  &&\!\!\!\!\!\!
  v_n^2[2] = \left\langle\frac
  {\int d\Delta\phi \,\cos(n\Delta\phi) \,\frac{dN_\mathrm{pairs}}
                                  {dy_1 dy_2 d\Delta\phi} 
          } 
  {\int d\Delta\phi \frac{dN_\mathrm{pairs}}
                                  {dy_1 dy_2 d\Delta\phi} 
          }\right\rangle,
\end{eqnarray}
where $\Delta\phi{\,=\,}\phi_1{-}\phi_2$ and the pair distribution has already been integrated over the average angle $\tilde\phi{\,\equiv\,}(\phi_1{+}\phi_2)/2$. 

Note that in Eqs.~(\ref{eq7}) the single-event averages are normalized by the number of pairs in the event, before averaging over events. This is important: Since the pair multiplicity fluctuates from event to event and within a multiplicity bin, and multiplicity anti-correlates with impact parameter with which the magnitudes of some of the anisotropic flow coefficients are geometrically correlated, this event-wise normalization avoids biasing the measured flow coefficients towards their values in events with larger than average multiplicity. 

Our definition of the integrated flow $v_n[2]$ agrees with the standard definition for the ``two-particle cumulant'' flow $v_n\{2\}$ \cite{Borghini:2000sa,Borghini:2001vi,Bilandzic:2010jr}, but the same is not true for the differential flow $v_n[2](\pT)$ which differs from $v_n\{2\}(\pT)$. The experimental definition of $v_n\{2\}(\pT)$ is \cite{Borghini:2000sa,Borghini:2001vi,Bilandzic:2010jr}
\begin{eqnarray}
\label{eq9a}
  &&v_n\{2\}(\pT) := \langle\{e^{in\phi_1}\}_{p_\mathrm{T1}} \{e^{-in\phi_2}\}\rangle/v_n\{2\}
 \nonumber\\
  && = \Bigl\langle v_n(\pT) v_n \cos[n(\Psi_n(\pT){-}\Psi_n)]\Bigr\rangle/v_n[2]\ .
\end{eqnarray}
Here only the first of the two particles within an event is taken from the desired $\pT$ bin and particle species; it is correlated with {\em all} other particles detected in the event, with obvious statistical advantages compared with $v_n[2](\pT)$ which requires both particles to be of the same kind and from the same $\pT$ bin. The normalization factor is the total rms flow of all charged hadrons. The last expression shows that $v_n\{2\}(\pT)$ reduces to $v_n[2](\pT){\,=\,}\sqrt{\langle v_n^2(\pT)\rangle}$ if and only if the flow angle $\Psi_n$ does not depend on $\pT$, the event-by-event fluctuations of $v_n(\pT)$ affect only its normalization but not the shape of its $\pT$ dependence, and the $v_n$ fluctuations of the particle species of interest are proportional to those of all hadrons. All of these assumptions are violated in hydrodynamic simulations of bumpy expanding fireballs. The difference between $v_n\{2\}(\pT)$ and $v_n[2](\pT)$ is thus sensitive to event-by-event fluctuations of the $\pT$-dependent difference $\Psi_n(\pT){-}\Psi_n$ between the flow angle of particles with momentum $\pT$ and the average event flow angle,\footnote{More precisely, $v_n\{2\}(\pT)$ depends on the difference between $\Psi_n(\pT)$ 
   {\em of the particles of interest} and the average flow angle $\Psi_n$ {\em of all detected 
   particles}. We checked numerically that the average hydrodynamic flow angles $\Psi_n$ 
   for identified pions and protons agree with great precision with the average flow angles for
   all particles in the event: Computing the ensemble average of 
   $\langle\cos[n(\Psi_n^\pi{-}\Psi_n^p)]\rangle$ for all harmonics $n$ and all collision 
   centralities, we found deviations of less than $1{-}2\%$ in all cases except for some of the 
   high-order harmonics with $n{\,>\,}6$ whose calculation is plagued by numerical errors at
   low $\pT{\,\lesssim\,}0.2$\,GeV arising from the finite grid spacing of our square numerical
   grid used in solving the hydrodynamic equations.} 
in addition to the (largely independent) fluctuations in the magnitudes of $v_n$ and $v_n(\pT)$.

Another approach to isolating effects arising from the $\pT$-dependence of the flow angles is a comparison of the $\pT$-dependent rms flow $v_n[2](\pT)$ with the so-called event-plane flow\footnote{One can replace the cosine function in this definition by the 
    exponential, omitting taking the real part in the second line, since the flow-angle fluctuations
    are symmetrically distributed such that the imaginary part vanishes after taking the event 
    average (this has been verified numerically).}$^,$\footnote{Note that we define the 
    $n^\mathrm{th}$-order event-plane flow relative to the $n^\mathrm{th}$-order flow 
    plane $\Psi_n$, and not relative to the elliptic flow plane $\Psi_2$ as sometimes done.}
\begin{eqnarray}
\label{eq10}
  &&v_n\{\EP\}(\pT) :=
  \left\la \frac{\int d\phi \, \cos[n(\phi{-}\Psi_n)]\,\frac{dN}{dy \pT d\pT d\phi}}
                    {\int d\phi \, \frac{dN}{dy \pT d\pT d\phi}} \right\ra
  \nonumber\\
  &&= \Re\Bigl\langle \{e^{in\phi}\}_{\pT}e^{-in\Psi_n}\Bigr\rangle 
  = \Re\Bigr\langle v_n(\pT) e^{in(\Psi_n(\pT){-}\Psi_n)}\Bigr\rangle
  \nonumber\\
  &&= \Bigl\langle v_n(\pT) \cos[n(\Psi_n(\pT){-}\Psi_n)]\Bigr\rangle .             
\end{eqnarray}
The equality in the second line arises from Eq.~(\ref{eq2}). Here for each event the ``average flow angle'' $\Psi_n$ is first obtained by computing the $\bm{Q}_n$ vector \cite{Poskanzer:1998yz}
\begin{equation}
\label{eq11}
  \bm{Q}_n = Q_n e^{in\Psi_n} := \frac{1}{N} \sum_{k=1}^N \omega_k\,e^{in\phi_k}
\end{equation}
(where $N$ is the number of detected particles in the event) and determining its phase. In 
principle, different choices for the weights $\omega_k$ can be considered \cite{Voloshin:2008dg}, but for consistency with Eq.~(\ref{eq10}) one must choose $\omega_k{\,=\,}1$. The ``average angle'' $\Psi_n$ for the event extracted from $\bm{Q}_n$ in general depends on the types of particles included in the sum in Eq.~(\ref{eq11}). As noted in footnote 2, however, the average flow angle for particles emitted from a hydrodynamic source is (within numerical precision) the same for all particle species, and the precision of extracting $\Psi_n$ in experiments can thus be maximized by including all detected particles in the $\bm{Q}_n$ vector (\ref{eq11}).\footnote{Since in this paper we ignore finite particle 
    statistical fluctuations in the final state, we know $\Psi_n$ with infinite precision for each
    particle species, and we will simply use these particle-specific values in our numerical 
    results below.}

The last line in Eq.~(\ref{eq10}) makes it clear that the differential event-plane flows $v_n\{\EP\}(\pT)$ are sensitive to the event-by-event fluctuations of the $\pT$-dependent flow angles $\Psi_n(\pT)$ around the ``average flow'' angle $\Psi_n$. Just like the finite number statistical fluctuations\footnote{Due
    to the finite number of particles detected in each event, the accuracy of determining 
    $\Psi_n$ is limited by finite number statistics, and an accurate experimental estimation 
    of the event-plane flow $v_n\{\EP\}$ requires an ``event-plane resolution correction'' 
    \cite{Voloshin:2008dg}. As shown in \cite{Alver:2008zza,Ollitrault:2009ie} (see also 
    the discussion in \cite{Gardim:2012im}), which moment of the underlying $v_n$ distribution 
    is actually measured by the total event-plane flow $v_n\{\EP\}$ depends on this 
    event-plane resolution: for perfect resolution $v_n\{\EP\}$ approaches the average flow 
    $\langle v_n\rangle$ whereas in the case of poor resolution it is closer to the rms flow 
    $v_n[2]{\,=\,}v_n\{2\}$ \cite{Alver:2008zza,Ollitrault:2009ie}. The mathematical analysis 
    in \cite{Ollitrault:2009ie} applies only to the integrated flow which allowed to ignore the 
    $\pT$-dependence of $v_n$ fluctuations as well as initial-state related, $\pT$-dependent 
    fluctuations of the flow angles that are not caused by finite multiplicity in the final state. 
    In view of the latter, event-plane resolution effects on differential flow measurements 
    and their correction require a new analysis.}
of the flow angle reconstructed from $\bm{Q}_n$ around the ``true'' flow angle of the event, these fluctuations smear out the azimuthal oscillations of the transverse momentum spectra and thus reduce the oscillation amplitudes $v_n\{\EP\}(\pT)$. In contrast to the former, they arise from fluctuations in the initial state and thus cannot be eliminated by improving or accounting for the resolution of the measurement of the final state. They carry valuable physical information about the initial state and the dynamics of its evolution into the final state.

We can remove the sensitivity of the measured quantity to the $\pT$-dependent fluctuations of the flow angle by first computing {\em for each event} the magnitude $v_n(\pT)$ of 
$\{e^{in\phi}\}_{\pT}{\,=\,}V_n(\pT)$, before summing over events:
\begin{eqnarray}
\label{eq12}
 && \langle v_n(\pT) \rangle =
       \left\langle\left\vert\{e^{in\phi}\}_{\pT}e^{-in\Psi_n}\right\vert\right\rangle 
       = \left\langle\left\vert\{e^{in\phi}\}_{\pT}\right\vert\right\rangle 
 \nonumber\\
 &&=\left\langle \sqrt{\{\cos(n\phi)\}^2_{\pT} +
                                \{\sin(n\phi)\}^2_{\pT}} \right\rangle                               
                               .\quad
\end{eqnarray}
Since the quantity inside the event average does not depend on the average flow angle $\Psi_n$, this observable is not subject to an event-plane resolution correction. However, due to finite multiplicity in the final state, the right hand side will still in general be positive and non-zero experimentally even if there is no underlying anisotropic flow in the event. Again, how to properly account for such finite sampling statistical effects requires additional analysis.  

By comparing $\langle v_n(\pT) \rangle$ (\ref{eq12}) with $v_n[2](\pT)$ (\ref{eq5},\ref{eq7}), $v_n\{2\}(\pT)$ (\ref{eq9a}), and $v_n\{\EP\}(\pT)$ (\ref{eq10}), we can experimentally assess and separate the relative importance of event-by-event fluctuations in the magnitudes and directions of the anisotropic flows as functions of $\pT$.

Let us now proceed to two-particle correlations between particles of different (but specified) momenta. Since in the first line of Eq.~(\ref{eq7}) both particles are taken from the same bin in $\pT$, the flow angle $\Psi_n(\pT)$ drops out from the expression. This is not true for azimuthal correlations between two particles with different $\pT$ \cite{Gardim:2012im}. In this case one finds \cite{Gardim:2012im,Luzum:2011mm} 
\begin{eqnarray}
\label{eq8}
  &&\tilde{V}_{n\Delta}(p_\mathrm{T1},p_\mathrm{T2}) 
  := \left\langle\{e^{in(\phi_1{-}\phi_2)}\}_{p_\mathrm{T1}p_\mathrm{T2}}\right\rangle
  \nonumber\\
  &&= \left\langle\{e^{in\phi_1}\}_{p_\mathrm{T1}}\{e^{-in\phi_2}\}_{p_\mathrm{T2}}\right\rangle
  = \left\langle V_n(p_\mathrm{T1}) V^*_n(p_\mathrm{T2})\right\rangle
  \nonumber\\
  &&=\left\langle v_n(p_\mathrm{T1})v_n(p_\mathrm{T2}) e^{in(\Psi_n(p_\mathrm{T1}){-}
  \Psi_n(p_\mathrm{T2}))}\right\rangle
  \nonumber\\
  &&=\Bigl\langle v_n(p_\mathrm{T1})v_n(p_\mathrm{T2}) \cos[n(\Psi_n(p_\mathrm{T1}){-}
  \Psi_n(p_\mathrm{T2}))]\Bigr\rangle.\quad\quad
\end{eqnarray}
Due to parity symmetry, $\tilde{V}_{n\Delta}(p_\mathrm{T1},p_\mathrm{T2})$ is real: while the quantity inside the event average $\langle\dots\rangle$ is in general complex for each individual event, its imaginary part averages to zero when summed over many events.

To properly account for multiplicity fluctuations, in Eq.\,(\ref{eq8}) the averages $\{\dots\}_{p_\mathrm{Ti}}$ within an event are once again normalized by the total number of particles included in the average, similar to Eq.~(\ref{eq7}). For this reason, $\tilde{V}_{n\Delta}(p_\mathrm{T1},p_\mathrm{T2}){\,=\,}\left\langle\{\cos(n\Delta\phi)\}_{p_\mathrm{T1},p_\mathrm{T2}}\right\rangle$ defined in Eq.~(\ref{eq8}) is not identical with the experimental quantity $V_{n\Delta}(p_\mathrm{T1},p_\mathrm{T2})$ which is obtained from a Fourier decomposition with respect to the difference angle $\Delta\phi$ of the two-particle distribution obtained by summing over many events, without normalizing the contribution from each event by the corresponding event multiplicity \cite{Alver:2010rt,Aamodt:2011by,Chatrchyan:2012wg,Luzum:2011mm,ATLAS:2012at}. For a meaningful comparison between theory and experiment, one should either normalize on the experimental side the contribution from each event to the two-particle distribution by the number of pairs in the event, or weight the theoretical prediction for $v_n(p_\mathrm{T1})v_n(p_\mathrm{T2}) e^{in(\Psi_n(p_\mathrm{T1}){-}\Psi_n(p_\mathrm{T2}))}$ for each event $i$ with a factor $N^{(i)}_{\mathrm{pairs}}/\langle N_\mathrm{pairs}\rangle$ before summing over events. We prefer the first option since it avoids the geometric bias arising from the correlation between collision geometry and particle multiplicity.

%
\begin{figure*}[ht]
  \begin{center}
    \includegraphics[width=0.7\linewidth]{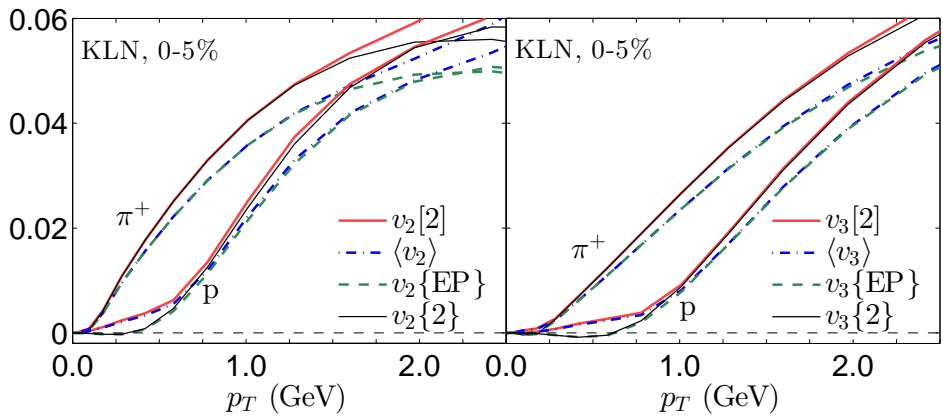}
  \end{center}
  \caption{(Color online) Comparison between the different definitions of the differential flows 
    $v_n[2](\pT)$ (\ref{eq5},\ref{eq7}), $v_n\{2\}(\pT)$ (\ref{eq9a}), 
    $v_n\{\EP\}(\pT)$ (\ref{eq10}), and $\langle v_n(\pT) \rangle$ (\ref{eq12}),  for pions 
    and protons from central ($0{-}5\%$ centrality) Pb+Pb collisions at 
    $\sqrt{s}{\,=\,}2.76\,A$\,TeV, computed with the viscous hydrodynamic code 
    {\tt VISH2{+}1}. See text for discussion.}
  \label{F2}
\end{figure*}
%
%
\begin{figure*}[ht]
  \begin{center}
    \includegraphics[width=0.7\linewidth]{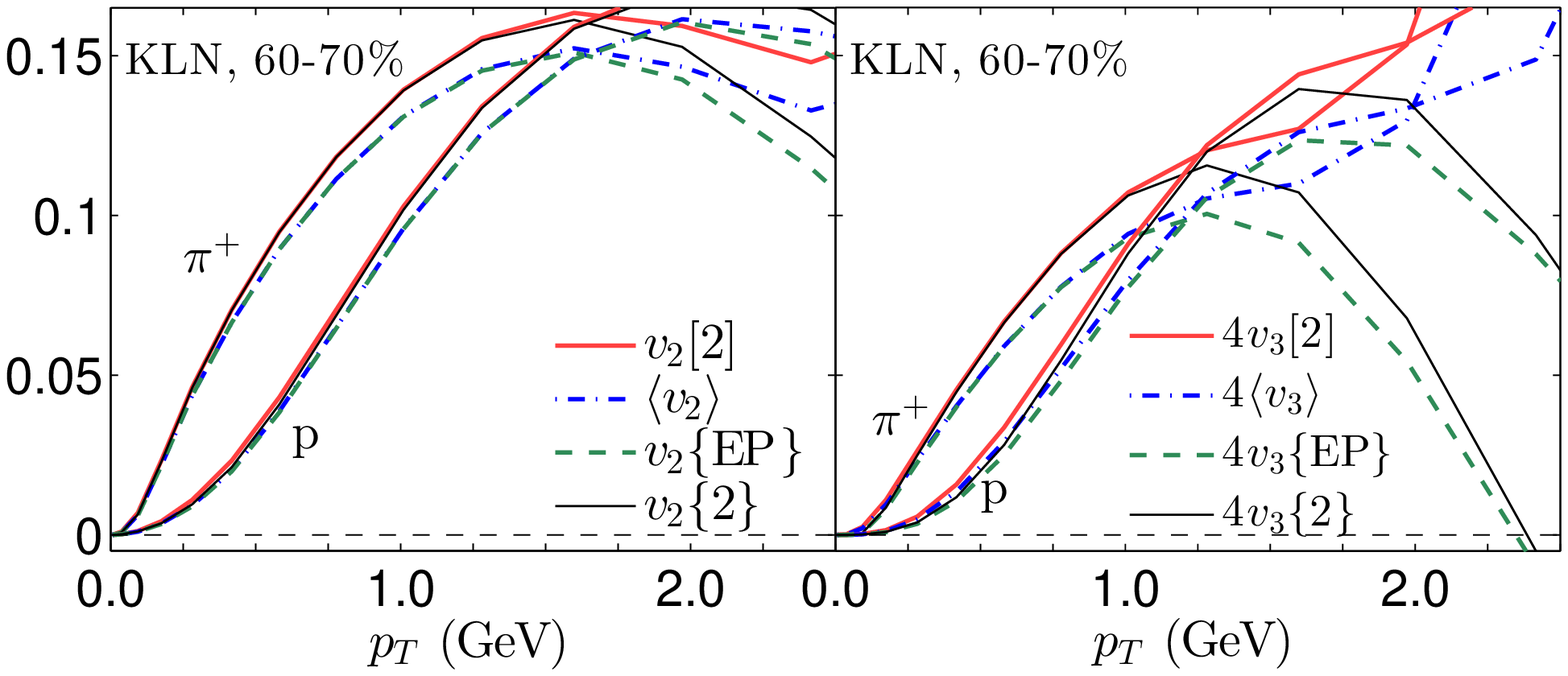}
  \end{center}
  \caption{(Color online) Same as Fig.~\ref{F2}, but for peripheral Pb+Pb collisions
      at $60{-}70\%$ centrality.}
  \label{F3}
\end{figure*}
%

\begin{figure*}[ht]
  \begin{center}
    \includegraphics[width=0.65\linewidth]{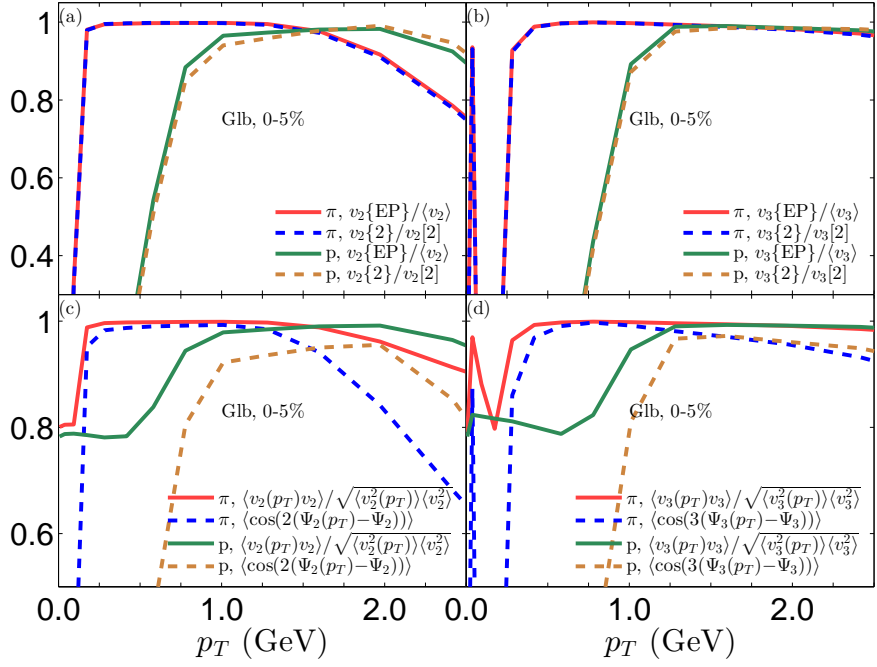}
  \end{center}
  \caption{(Color online) (a,b) Ratios of differently defined elliptic (a) and triangular (b) flow 
     coefficients for pions and protons as functions of $\pT$. (c,d) $\pT$-dependence of the
     separate fluctuations of the magnitudes $v_n$ and the angles $\Psi_n$ for pions and 
     protons as discussed in the text. All curves from viscous hydrodynamics with 
     $\eta/s{\,=\,}0.08$ for central 2.76\,$A$\,TeV Pb+Pb collisions with MC-Glauber initial
     conditions. Results for MC-KLN initial conditions evolved with $\eta/s{\,=\,}0.2$ look very 
     similar.
     }
  \label{F4}
\end{figure*}

\begin{figure*}[ht]
  \begin{center}
    \includegraphics[width=0.63\linewidth]{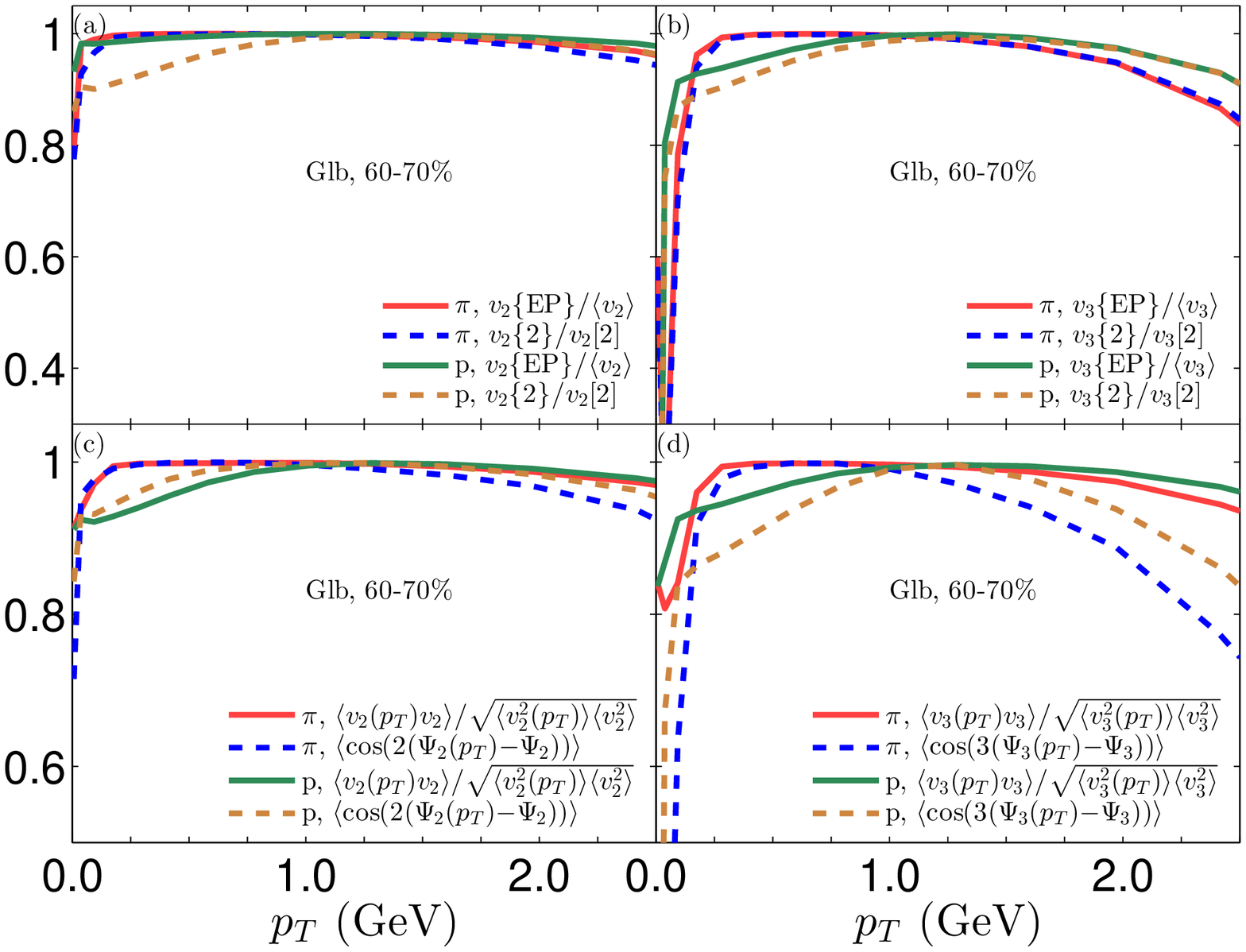}
  \end{center}
  \caption{(Color online) Same as Fig.~\ref{F4}, but for peripheral collisions at 
     $60{-}70\%$ centrality.
        }
  \label{F5}
\end{figure*}

\begin{figure*}[ht]
  \begin{center}
    \includegraphics[width=\linewidth]{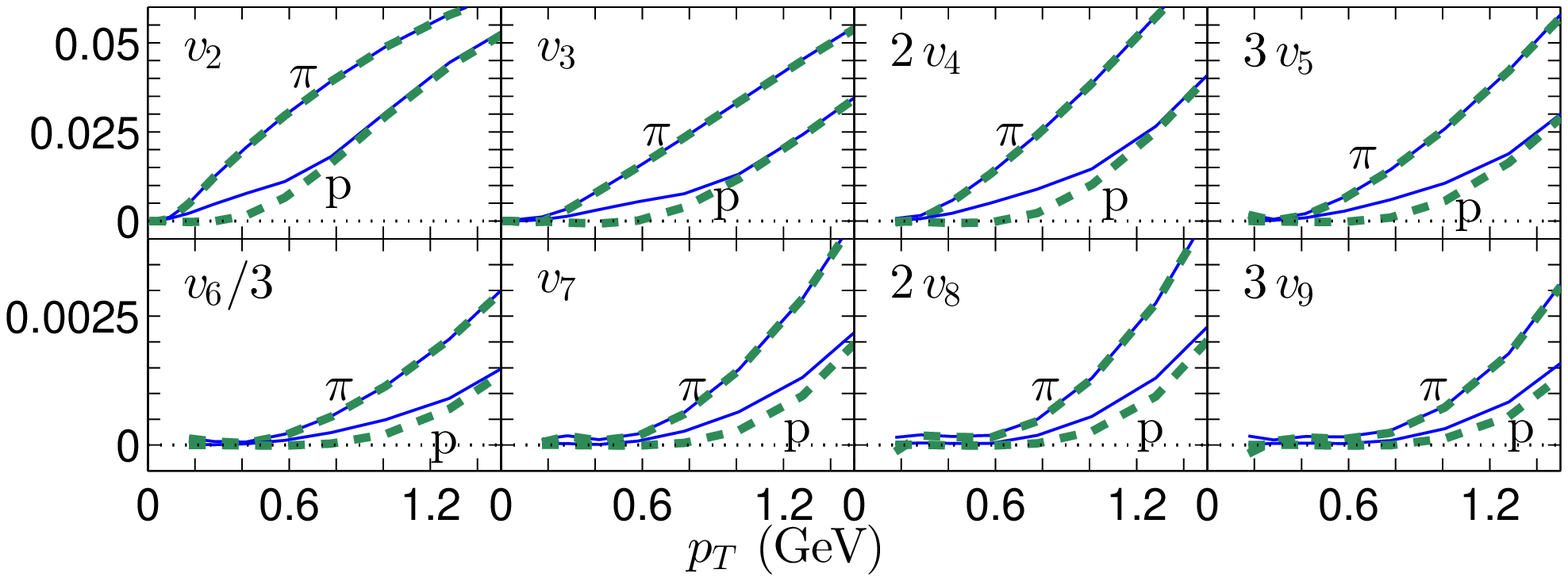}
  \end{center}
  \caption{(Color online) Similar to Figs.~\ref{F2},\ref{F3}, but with MC-Glauber initial conditions
    for Pb+Pb collisions in the $0{-}10\%$ centrality bin. For clarity only $v_n\{\EP\}(\pT)$ 
    (Eq.~(\ref{eq10}), thick dashed) and $\langle v_n(\pT) \rangle$ (Eq.~(\ref{eq12}), thin solid)
    are shown, but for all harmonics from $n{\,=\,}2$ to $n{\,=\,}9$ (scaled by appropriate factors 
    for best visibility). This set of plots focusses on the low-$\pT$ region $\pT{\,<\,}1.5$\,GeV 
    where the effects from flow angle fluctuations are strongest. See text for discussion.}
  \label{F6}
\end{figure*}

Equation~(\ref{eq8}) makes it obvious that the two-particle correlation coefficient $\tilde{V}_{n\Delta}(p_\mathrm{T1},p_\mathrm{T2})$ does not factorize into a product of single-particle anisotropic flow coefficients \cite{Gardim:2012im}. There are two contributions to this breaking of factorization: $\pT$-dependent event-by-event fluctuations of the magnitude of the flow coefficient $v_n$, and $\pT$-dependence of the flow angles $\Psi_n$ \cite{Gardim:2012im} (which also fluctuate from event to event). It is possible to define a non-factorizing correlator that is only affected by the fluctuations of $v_n(\pT)$ but insensitive to the flow angles:
\begin{eqnarray}
\label{eq9}
  &&\Bigl\langle  v_n(p_\mathrm{T1})v_n(p_\mathrm{T2})\Bigr\rangle 
  \nonumber\\
  &&=
  \left\langle \sqrt{\{\cos(n\Delta\phi)\}^2_{p_\mathrm{T1},p_\mathrm{T2}} +
                            \{\sin(n\Delta\phi)\}^2_{p_\mathrm{T1},p_\mathrm{T2}}}\right\rangle.\qquad
\end{eqnarray}
It is obtained experimentally by first obtaining the magnitude of the quantity $\{e^{in\phi_1}\}_{p_\mathrm{T1}}\{e^{-in\phi_2}\}_{p_\mathrm{T2}}$ for each event, normalizing it to the number of pairs used for its computation, and than adding the results for many events. Its sensitivity to finite number statistical effects should be similar to Eq.~(\ref{eq12}) and needs to be explored. By comparing the quantity $\tilde{V}_{n\Delta}(p_\mathrm{T1},p_\mathrm{T2})$ from Eq.~(\ref{eq8}) with $\langle  v_n(p_\mathrm{T1})v_n(p_\mathrm{T2})\rangle$ from Eq.~(\ref{eq9}) one can assess the importance of the $\pT$-dependence and event-by-event fluctuations of the flow angles $\Psi_n$ (which affect the former but not the latter).

\section{The effect of flow fluctuations on differential $\bm{v_n}$ measures}
\label{sec3}

In this section we compare the differential flows $v_n(\pT)$ extracted from the 22,000 viscous hydrodynamic simulations per centrality bin of 2.76\,$A$\,TeV Pb+Pb collisions at the LHC (11,000 each with MC-Glauber and MC-KLN initial density profiles) that were generated in Ref.~\cite{Qiu:2012uy}. We use the Cooper-Frye prescription to compute from the hydrodynamic output on the freeze-out surface the single-particle distributions $dN/(dy\pT d\pT d\phi)$ as continuous functions of $\pT$ and $\phi$ (i.e. we do not sample the distribution to generate a finite number of particles per event, but pretend that the spectrum is sampled infinitely finely -- this avoids the need to correct for effects arising from finite number statistics, such as imperfect event-plane resolution). All resonance decays are included in the final stable hadron spectra. The details of the hydrodynamic simulations, initial conditions and freeze-out parameters are not important for the qualitative study presented here, but the interested reader can find them described in Refs.~\cite{Qiu:2012uy,Qiu:2011hf,Shen:2011eg}. Here we only note that MC-Glauber (MC-KLN) initial conditions were hydrodynamically evolved with specific shear viscosity $\eta/s{\,=\,}0.08$ (0.2). 

We present results for pions and protons, representing light and heavy particle species. Qualitatively, although not quantitatively, the same generic features are observed with MC-KLN and MC-Glauber model initial density profiles, and we show examples of both. Figures ~\ref{F2} and \ref{F3} show elliptic and triangular flows in their left and right panels, for central (Fig.~\ref{F2}) and peripheral (Fig.~\ref{F3}) Pb+Pb collisions. The reader should compare the curves for $v_{2,3}[2]$ and $\langle v_{2,3}\rangle$, which are not affected by flow angle fluctuations ({\it c.f.} Eqs.~(\ref{eq5}) and (\ref{eq12})), with those for $v_{2,3}\{2\}$ and $v_{2,3}\{\EP\}$, which {\em are} affected by the $\pT$-dependence of the flow angles $\Psi_n$ and their event-by-event fluctuations ({\it c.f.} Eqs.~(\ref{eq9a}) and (\ref{eq10})): For protons with transverse momenta below about 1\,GeV, flow angle fluctuations are seen to cause a significant suppression of the latter (in some cases even leading to negative elliptic flow values).\footnote{Note that the factor $\cos[n(\Psi_n(\pT){-}\Psi_n)]$
   in Eqs.~(\ref{eq9a}) and (\ref{eq10}) is maximal if $\Psi_n(\pT)$ is always aligned with 
   $\Psi_n$. The suppression of, say, $v_n\{\EP\}(\pT)$ relative to $\langle v_n \rangle(\pT)$
   does therefore not indicate a definite momentum tilt of the emitting source at a given 
   $\pT$ relative to the average $\Psi_n$, but simply reflects a nonzero difference 
   $\Psi_n(\pT){-}\Psi_n$ that fluctuates from event to event, suppressing the value of  
   $\cos[n(\Psi_n(\pT){-}\Psi_n)]$ for either sign of the difference.}
For the much lighter pions flow angle fluctuation effects are almost invisible at low $\pT$. For protons they gradually disappear, too, as one goes from central (Fig.~\ref{F2}) to peripheral (Fig.~\ref{F3}) collisions.\footnote{The curves shown in Figs.~\ref{F2},\,\ref{F3} include the
   decay products from unstable hadronic resonances. We have observed that for protons 
   the flow angle fluctuation induced difference at low $\pT$ between ($v_{2,3}[2]$, 
   $\langle v_{2,3}\rangle$) on the one hand and ($v_{2,3}\{2\}$, $v_{2,3}\{\EP\}$) on the other 
   hand doubles if only directly emitted (``thermal'') particles are included in the analysis. 
   Resonance decays thus dilute the sensitivity of the proposed observables to flow angle 
   fluctuations by about 50\%.}

Event-by-event fluctuations of the {\em magnitudes} of $v_{2,3}$ are accessible by comparing $\langle v_{2,3}\rangle$ with $v_{2,3}[2]{\,=\,}\langle v^2_{2,3}\rangle^{1/2}$. When plotting the ratios $v_{2,3}[2](\pT)/\langle v_{2,3}\rangle(\pT)$ for central ($0{-}5\%$ centrality) collisions, where anisotropic flows are caused exclusively by fluctuations, with negligible geometric bias from a non-zero average deformation of the nuclear overlap region, we found for both pions and protons a constant (i.e. $\pT$-independent) value of $2/\sqrt{\pi}{\,\approx\,}1.13$. This is expected \cite{Voloshin:2007pc,Luzum:2012da}: If the flow angle $\Psi_n$ is randomly distributed relative to the reaction plane, the components of $V_n(\pT)$ along and perpendicular to the reaction plane are approximately Gaussian distributed around zero, and the magnitude $v_n(\pT)$ of the complex flow coefficient is Bessel-Gaussian distributed with $\sqrt{\langle v_n^2(\pT)\rangle}{\,=\,}\frac{2}{\sqrt{\pi}}\langle v_n(\pT)\rangle$ (see Eqs.\,(4) and (5) in Ref.~\cite{Voloshin:2007pc}). A similar $\pT$-independent ratio is not observed at larger impact parameters: even for triangular flow, which continues to be fluctuation-dominated also at non-zero impact parameters, we observe deviations of the ratio $v_3[2](\pT)/\langle v_3\rangle(\pT)$ from $2/\sqrt{\pi}$ at both low and high $\pT$; for elliptic flow these deviations are larger and significant at all $\pT$.  

Interestingly, for central collisions we found approximately the same constant value $2/\sqrt{\pi}$ for the ratio $v_{2,3}\{2\}(\pT)/v_{2,3}\{\EP\}(\pT)$ (except near the $\pT$ values where either the numerator or denominator passes through zero). Looking at the definitions (\ref{eq9a}) and (\ref{eq10}), this suggests an approximate factorization of the $\pT$-dependent flow angle fluctuations (which enter through the factor $\cos[n(\Psi_n(\pT){-}\Psi_n)]$ that cancels between numerator and denominator if it fluctuates independently) from the fluctuations of the magnitude $v_n(\pT)$, as well as an approximate $\pT$-independence of the $v_{2,3}$ fluctuations.

To follow up on these observations and gain deeper insight into the relative importance of flow angle fluctuations in different $\pT$ ranges, let us look at Figs.~\ref{F2},\,\ref{F3} and note that the frequently measured quantity $v_{2,3}\{2\}(\pT)$ behaves like the event-plane flow $v_{2,3}\{\EP\}(\pT)$ at low $\pT$ and like the differential rms flow $v_{2,3}[2](\pT)$ at intermediate $\pT$. This suggests that it is dominated by flow angle fluctuations at low $\pT$ and by fluctuations of the magnitude of $v_{2,3}(\pT)$ at higher $\pT$). In central collisions, the proton $v_{2,3}\{2\}(\pT)$ even turns negative at low $\pT$, whereas $v_{2,3}[2](\pT)$ is by definition always positive. A related observation is that the proton event-plane flow $v^p_{2,3}\{\EP\}(\pT)$ in Fig.~\ref{F2} approximately agrees with $v^p_{2,3}\{2\}(\pT)$ at low $\pT$ (where flow angle fluctuations seem to have strong effects) but with the mean flow $\langle v^p_{2,3}(\pT)\rangle$ at higher $\pT$ (where flow angle fluctuation effects are weak). This is reminiscent of the behavior of the {\em $\pT$-integrated} event-plane flow which approaches the mean flow for good event-plane resolution. Flow angle fluctuations appear to have similar effects on flow measures as a decrease in flow angle resolution. The difference is that the former is a physical effect due to initial-state fluctuations whereas the latter is a finite sampling statistical effect in the final state and affected by detector performance.

To make these qualitative observations quantitative, we plot in the upper two panels of Fig.~\ref{F4} the ratios $v_{2,3}\{2\}/v_{2,3}[2]$ and $v_{2,3}\{\EP\}/\langle v_{2,3}\rangle$ as functions of $\pT$, for both pions and protons. (We focus here on the results from Fig.~\ref{F2} for central collisions where all anisotropic flows are fluctuation-dominated.) In each case the numerator is sensitive to the flow angle fluctuations while the denominator is not. However, numerator and denominator are also differently affected by fluctuations in the magnitudes of $v_n$. Both ratios are seen to behave very similarly, staying close to 1 at intermediate $p_T$ but dropping steeply at low $\pT$ and more moderately at high $\pT$. The steep drop at low transverse momenta sets in at $\pT{\,\sim\,}1$\,GeV for protons, but at much smaller $\pT{\,<\,}0.25$\,GeV for pions. We do not have a full understanding of this mass dependence, beyond the qualitative observation that the minimum of the variance of the flow angle fluctuations shown in Figs.~\ref{F1}b,c is shifted to higher $\pT$ for protons compared to pions, and that quite generally strong radial flow shifts all flow anisotropies to higher $\pT$ values for heavier particles.

The lower two panels of Fig.~\ref{F4} demonstrate that the behavior of the ratios shown in the two upper panels is strongly dominated by flow angle fluctuations. The dashed lines in Figs.~\ref{F4}c,d show the flow angle fluctuations $\langle\cos[n(\Psi_n(\pT){-}\Psi_n)]\rangle$ in isolation. Their $\pT$ dependence alone is almost sufficient to completely explain the shape of the curves
in panels (a) and (b). The solid lines in Figs.~\ref{F4}c,d show that at intermediate $\pT$ fluctuations in the magnitudes of the $\pT$-dependent flow $v_n(\pT)$ and the $\pT$-integrated $v_n$ tend to be correlated with each other ($v_n(\pT)\propto v_n$) while they appear to fluctuate more independently at low and high $\pT$. At high $\pT$ this decorrelation contributes to the suppression of the ratios shown in panels (a,b). At low $\pT$, the decorrelation of the $\pT$-dependent flow {\em magnitude} fluctuations $v_n(\pT)$ from the $\pT$-integrated flow $v_n$ does not become effective until after the ratios have already been suppressed by flow {\em angle} fluctuations, and its effect is therefore subdominant. 

%
\begin{figure*}
  \includegraphics[width=0.8\linewidth]{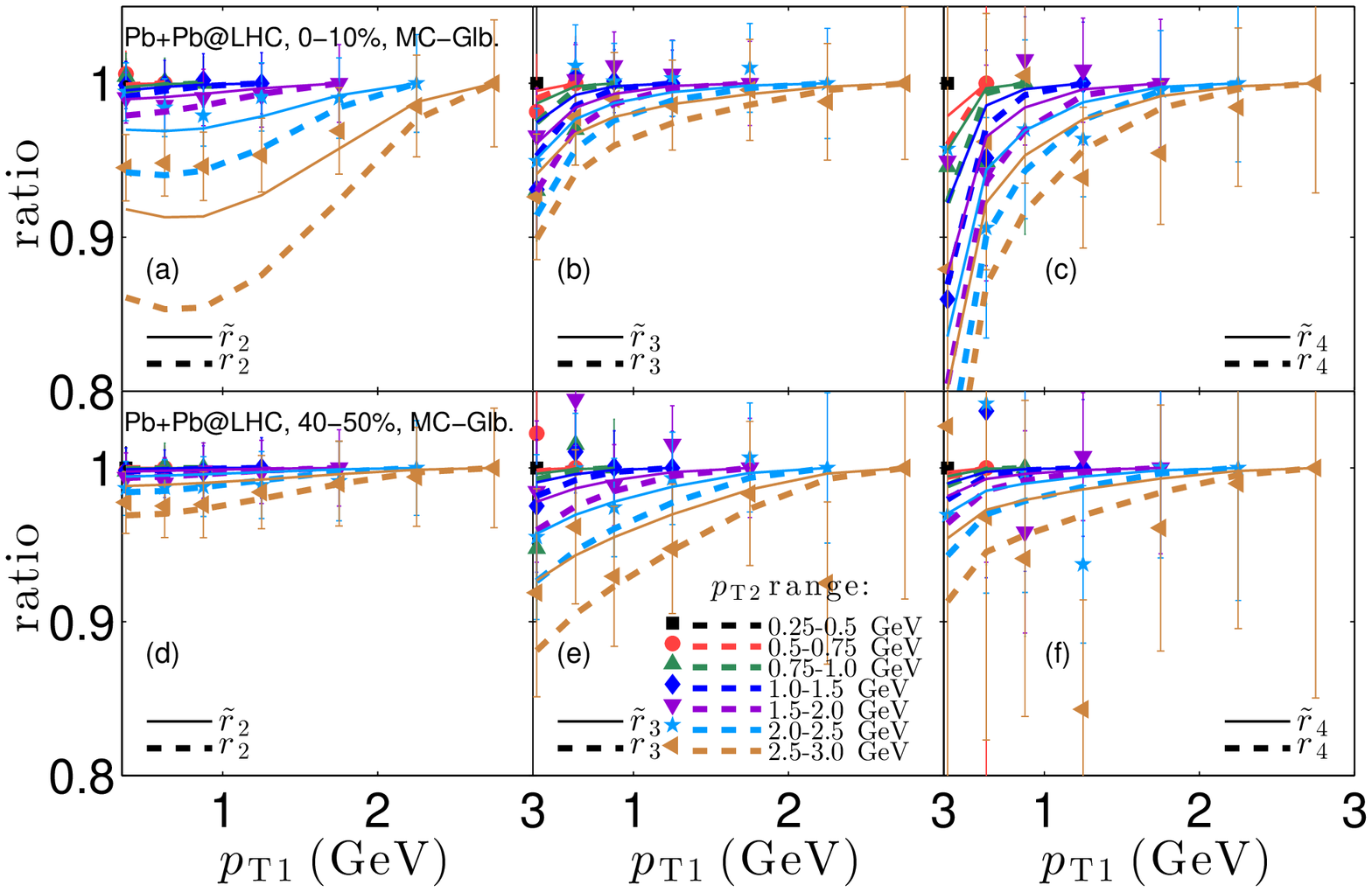}
  \caption{(Color online) The ratios $r_{2,3,4}(p_\mathrm{T1},p_\mathrm{T2})$ (thick dashed 
      lines) and $\tilde{r}_{2,3,4}(p_\mathrm{T1},p_\mathrm{T2})$ (thin solid lines), defined in
      Eqs.~(\ref{eq15}) and (\ref{eq16}), as functions of $p_\mathrm{T1}$ for different 
      $p_\mathrm{T2}$ ranges, as indicated. Filled symbols represent ALICE data for
      Pb-Pb collisions at $\sqrt{s}{\,=\,}2.76\,A$\,TeV \cite{Aamodt:2011by}. The lines are the
      corresponding viscous hydrodynamic calculations with MC-Glauber initial conditions, 
      using specific shear viscosity $\eta/s{\,=\,}0.08$. Panels (a,b,c) show $r_{2,3,4}$ and 
      $\tilde{r}_{2,3,4}$ for central ($0{-}10\%$) collisions, panels (d,e,f) show the same ratios 
      for peripheral ($40{-}50\%$) collisions.}
\label{F7}
\end{figure*}
%

In summary, we see for central collisions that at low $\pT$ the differences between $v_n\{2\}(\pT)$ and $v_n[2](\pT)$, as well as between $v_n\{\EP\}(\pT)$ and $\langle v_n(\pT)\rangle$, are dominated by flow angle fluctuations, whereas at high $\pT$ fluctuations of both the flow angles and flow magnitudes must be considered to explain their differences. At intermediate $\pT$ flow angle fluctuations appear to be unimportant, $v_n(\pT)$ fluctuates in sync with the $\pT$-integrated $v_n$, and the differences between $v_n\{2\}(\pT)$ and $v_n[2](\pT)$, as well as between $v_n\{\EP\}(\pT)$ and $\langle v_n(\pT)\rangle$, vanish.

Figure~\ref{F5} shows the same ratios as Fig.~\ref{F4} for peripheral Pb+Pb collisions, again using MC-Glauber initial conditions with $\eta/s{\,=\,}0.08$.\footnote{The main difference 
    with results from MC-KLN initial conditions with $\eta/s{\,=\,}0.2$ (not shown) is that the 
    latter exhibit stronger suppression effects from the flow fluctuation factor 
    $\cos[n(\Psi_n(\pT){-}\Psi_n)]$ in the high-$\pT$ region $\pT{\,\gtrsim\,}1$\,GeV (see 
    also Fig.~\ref{F3}).} 
Compared to central collisions (shown in Fig.~\ref{F4}), the flow angle fluctuation effects at low $\pT$ are much weaker and appear to be shifted to lower transverse momenta, for both pions and protons. At high $\pT{\,\gtrsim\,}1$\,GeV, Figs.~\ref{F5}c,d show that effects from fluctuations of the flow angles (dashed lines) dominate over those from fluctuations of the flow magnitudes (solid lines).

Finally, in Figure~\ref{F6} we explore (for near-central collisions) how the flow angle fluctuation effects, which push the event-plane flow $v_n\{\EP\}$ at low-$\pT$ below the value of the average flow $\langle v_n\rangle$, evolve as the harmonic order $n$ increases. (For $n{\,\geq\,}4$ we do not show results below $\pT{\,=\,}0.2$\,GeV, for technical reasons explained in footnote 2.) For pions, flow angle fluctuations are invisible in the shown $\pT$ region for all flow harmonics; for protons, they are clearly visible for all harmonic flows. The relative magnitude of their effect on the difference $\langle v_n\rangle(\pT){-}v_n\{\EP\}(\pT)$ at any fixed $\pT$ decreases as $n$ increases, but the difference remains nonzero over a larger $\pT$ range for the higher harmonics. 

\section{Non-factorization of flow-induced two-particle correlations}
\label{sec4}

\begin{figure*}
  \includegraphics[width=0.8\linewidth]{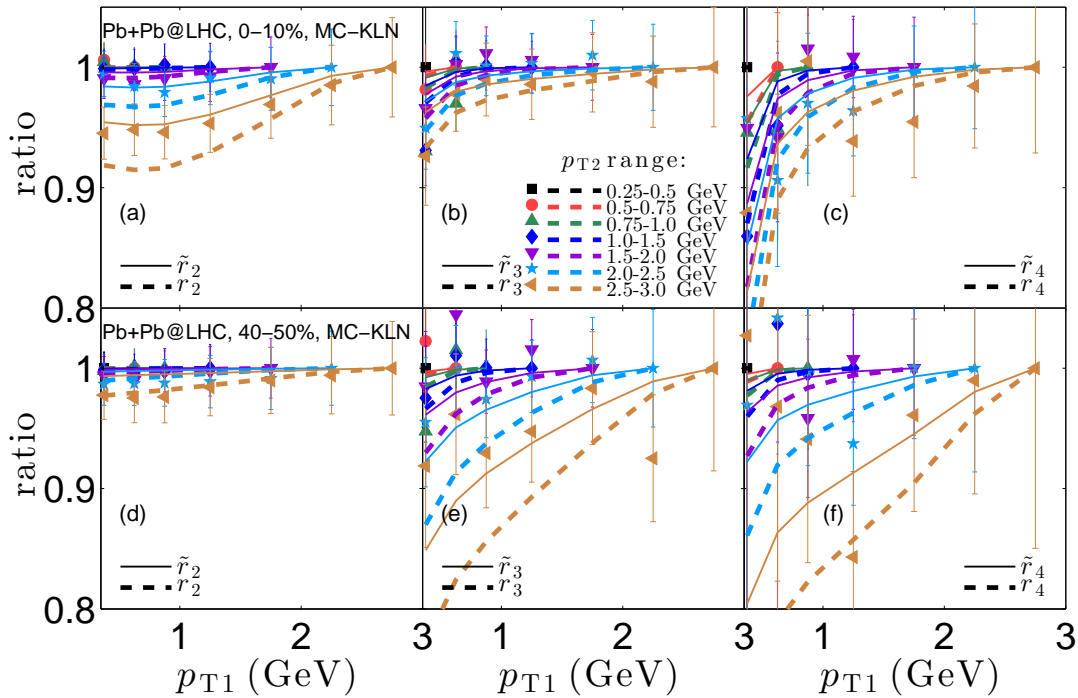}
  \caption{(Color online) Same as Fig.~\ref{F6}, but for MC-KLN initial conditions that have 
     been evolved hydrodynamically with $\eta/s{\,=\,}0.2$ (a shear viscosity value that is 2.5
     times larger than the one used in Fig.~\ref{F6}).}
\label{F8}
\end{figure*}

The breaking of factorization of flow-induced two-particle correlations by flow fluctuations was first emphasized by Gardim {\it et al.} \cite{Gardim:2012im}. Their study was based on simulations using ideal fluid dynamics, which are here repeated with viscous fluid dynamics. A comparison of Figs.~\ref{F7},\,\ref{F8} below with the plots shown in Ref.~\cite{Gardim:2012im} shows that viscous effects reduce the amount by which event-by-event fluctuations break factorization. We here explore the relative role played in this context by fluctuations in the magnitudes and angles of the flows.

%
%

To this end we define the following two ratios, both symmetric in $p_\mathrm{T1}$ and $p_\mathrm{T2}$:
\begin{eqnarray}
\label{eq15}
  &&r_n(p_\mathrm{T1},p_\mathrm{T2}) := 
  \frac{\tilde{V}_{n\Delta}(p_\mathrm{T1},p_\mathrm{T2})}
         {\sqrt{\tilde{V}_{n\Delta}(p_\mathrm{T1},p_\mathrm{T1})
                   \tilde{V}_{n\Delta}(p_\mathrm{T2},p_\mathrm{T2})}}
 \nonumber\\  
  && = \frac{\la V_n(p_\mathrm{T1}) V_n^*(p_\mathrm{T2}) \ra} 
                {\sqrt{\la |V_n(p_\mathrm{T1})|^2\ra \la |V_n(p_\mathrm{T2})|^2 \ra}}
 \\\nonumber 
 && = \frac{\langle v_n(p_\mathrm{T1})v_n(p_\mathrm{T2}) 
                 \cos[n(\Psi_n(p_\mathrm{T1}){-}\Psi_n(p_\mathrm{T2}))]\rangle}
                {\sqrt{\la v_n^2(p_\mathrm{T1})\ra\la v_n^2(p_\mathrm{T2})\ra}}; 
 \\\label{eq16}
  &&\tilde{r}_n(p_\mathrm{T1},p_\mathrm{T2}) 
 \\\nonumber
  && :=               
        \frac{\langle v_n(p_\mathrm{T1})v_n(p_\mathrm{T2}) 
                 \cos[n(\Psi_n(p_\mathrm{T1}){-}\Psi_n(p_\mathrm{T2}))]\rangle}
                {\la v_n(p_\mathrm{T1}) v_n(p_\mathrm{T2})\ra}. 
\end{eqnarray}
The ratio $r_n$, first introduced and studied with ideal fluid dynamics in \cite{Gardim:2012im}, is sensitive to fluctuations of both the magnitudes $v_n(\pT)$ and angles $\Psi_n(\pT)$ of the complex anisotropic flow coefficients $V_n(\pT)$ defined in Eq.~\ref{eq2}). The second ratio $\tilde{r}_n$, on the other hand, differs from unity only on account of flow angle fluctuations. By comparing the two ratios with each other and with experimental data we can isolate the role played by flow angle fluctuations in the breaking of factorization of the event-averaged two-particle cross section. In the absence of non-flow correlations both ratios are always ${\leq\,}1$.  

Figures~\ref{F7} show these ratios for all charged hadrons as functions of $p_\mathrm{T1}{\,\leq\,}p_\mathrm{T2}$ for fixed ranges of $p_\mathrm{T2}$, indicated by different colors.\footnote{The $p_\mathrm{T2}$ ranges are adjusted to the experimental  
    data, and the ratios were computed by first averaging the numerator and denominator 
    over the given $p_\mathrm{T2}$ range.}
Figs.~\ref{F7}a,b,c focus on central, Figs.~\ref{F7}d,e,f on peripheral collisions; in both cases, we used MC-Glauber initial conditions and evolved them with {\tt VISH2{+}1} using $\eta/s{\,=\,}0.08$ for the specific shear viscosity. In central collisions the hydrodynamic simulations appear to overpredict the factorization breaking effects, while in peripheral collisions theory and data agree somewhat better. More precise experimental data would be desirable. The comparison of $r_n$ (dashed lines) with $\tilde{r}_n$ shows that a significant fraction (${\,\sim\,}50\%$ or more) of the effects that cause the breaking of factorization arises from flow angle fluctuations. This seems to hold at all the transverse momenta shown in the figures. A comparison of the top and bottom rows of panels in Fig.~\ref{F7} shows that factorization-breaking effects are stronger for harmonics that are fluctuation dominated ({\it i.e.} all harmonics in central collisions, and the odd harmonics (especially $v_3$) in peripheral collisions) and appear to weaken for $v_2$ and $v_4$ in peripheral collisions where both the magnitudes $v_{2,4}$ and the flow angles
$\Psi_{2,4}$ are mostly controlled by collision geometry.

To explore the effects of shear viscosity of the expanding fluid on the breaking of factorization we show in Figure~\ref{F8} the same data as in Fig.~\ref{F7}, but compared with hydrodynamic calculations that use MC-KLN initial conditions evolved with $\eta/s{\,=\,}0.2$ (a 2.5 times larger viscosity than used in Fig.~\ref{F7}). Obviously, the MC-KLN model produces a different initial fluctuation spectrum than the MC-Glauber model, so not all of the differences between Figs.~\ref{F7} and \ref{F8} can be attributed to the larger viscosity. However, in conjunction with the ideal fluid results reported in \cite{Gardim:2012im}, the comparison of these two figures strengthens the conclusion that increased shear viscosity tends to weaken the fluctuation effects that cause the event-averaged two-particle cross section to no longer factorize.  

\section{Summary}
\label{sec5}

All experimental precision measures of anisotropic flow in relativistic heavy-ion collisions are based on observables that average over many collision events. It has been known for a while that both the magnitudes $v_n$ and flow angles $\Psi_n$ of the complex anisotropic flow coefficients $V_n$ fluctuate from event to event, but only very recently it became clear that not only the $v_n$, but also their associated angles $\Psi_n$ depend on $\pT$, and that the difference $\Psi_n(\pT){-}\Psi_n$ between the $\pT$-dependent and $\pT$-averaged flow angles also fluctuates from event to event. In the present study we have pointed out that these flow angle fluctuations leave measurable traces in experimental observables from which the ensemble-averaged $\pT$-dependent anisotropic flows are extracted. We have introduced several new flow measures and shown how their comparison with each other and with flow measures that are already in wide use allows to separately assess the importance of event-by-event fluctuations of the magnitudes and angles of $V_n{\,=\,}v_n e^{in\Psi_n}$ on  experimentally determined flow coefficients. 

Viscous hydrodynamic simulations show that flow angle fluctuations affect the $\pT$-dependent flow coefficients of heavy hadrons (such as protons) more visibly than those of light hadrons (pions). In near-central collisions, where anisotropic flow is dominated by initial density fluctuations rather than overlap geometry, the effects from flow angle fluctuations appear to be strongest for particles with transverse momenta $\pT{\,\lesssim\,}m$. A precise measurement and comparison of $\la v_n(\pT)\rangle$ (Eq. (\ref{eq12})), $v_n\{\EP\}(\pT)$ (Eq.~(\ref{eq10})), $v_n[2](\pT)$ (Eqs.~(\ref{eq5},\ref{eq7})), and $v_n\{2\}(\pT)$ (Eq.~(\ref{eq9a}) for identified pions, kaons and protons with transverse momenta $\pT{\,<\,}2$\,GeV should be performed to confirm the hydrodynamically predicted effects from flow angle fluctuations. The theoretical interpretation of these measurements requires a reanalysis of finite sampling statistical effects on the {\em $\pT$-dependent differential} flows, stemming from the finite multiplicity of particles of interest in a single event, which we did not consider here. The proposed comparison holds the promise of yielding valuable experimental information to help constrain the distribution of initial density fluctuations in relativistic heavy ion collisions and may prove crucial for a precision determination of the QGP shear viscosity. 

We also showed that flow angle fluctuations are responsible for more than half of the hydrodynamically predicted factorization breaking effects studied in Ref.~\cite{Gardim:2012im} and in Sec.~\ref{sec4} above, and that these effects are directly sensitive to the shear viscosity of the expanding fluid, decreasing with increasing viscosity. By combining the study of various types of differential anisotropic flow measures with an investigation of the flow-induced breaking of the factorization of two-particle observables into products of single-particle observables one can hope to independently constrain the fluid's transport coefficients and the initial-state fluctuation spectrum.


\acknowledgments{The authors thank Matt Luzum, Jean-Yves Ollitrault, Art Poskanzer, Raimond Snellings, and Sergei Voloshin for very illuminating and clarifying discussions that helped to sharpen the arguments presented here. This work was supported by the U.S.\ Department of Energy under Grants No.~\rm{DE-SC0004286} and (within the framework of the JET Collaboration) \rm{DE-SC0004104}.
}


\end{document}